\documentclass[reprint,
 superscriptaddress,
 bibnotes,
 amsmath,amssymb, aps,
 prl,
floatfix,
]{revtex4-1}

\usepackage[utf8]{inputenc}
\usepackage{graphicx}
\usepackage[caption=false]{subfig}
\usepackage{dcolumn}
\usepackage{multirow}
\usepackage{tabularx}
\usepackage{bm}
\usepackage{hyperref}
\usepackage{amsmath}
\usepackage{amssymb}
\usepackage{comment}
\usepackage{color}
\usepackage{enumerate}
\usepackage{booktabs}
\usepackage{siunitx}

\newcommand{\GeV}{\text{\,GeV}}

\newcommand{\half}{\frac{1}{2}}

\def\apj{\ref@jnl{ApJ}}                 

\newcolumntype{L}[1]{>{\hsize=#1\hsize\raggedright\arraybackslash}X}%
\newcolumntype{R}[1]{>{\hsize=#1\hsize\raggedleft\arraybackslash}X}%
\newcolumntype{C}[1]{>{\hsize=#1\hsize\centering\arraybackslash}X}%

\makeatletter
\newcommand{\thickhline}{%
    \noalign {\ifnum 0=`}\fi \hrule height 2pt
    \futurelet \reserved@a \@xhline
}
\newcolumntype{"}{@{\hskip\tabcolsep\vrule width 2pt\hskip\tabcolsep}}
\makeatother

\definecolor{deepblue}{rgb}{0,0,0.7}

\begin{document}

\title{Search for light scalar dark matter with atomic gravitational wave detectors}

\author{Asimina Arvanitaki}
\email{aarvanitaki@perimeterinstitute.ca}
\affiliation{Perimeter Institute for Theoretical Physics, Waterloo, Ontario N2L 2Y5, Canada}
\author{Peter W. Graham} 
\email{pwgraham@stanford.edu}
\affiliation{Stanford Institute for Theoretical Physics, Stanford University, Stanford, California 94305, USA}
\author{Jason M. Hogan} 
\email{hogan@stanford.edu}
\affiliation{Department of Physics, Stanford University, Stanford, California 94305}
\author{Surjeet Rajendran}
\email{surjeet@berkeley.edu}
\affiliation{Berkeley Center for Theoretical Physics, Department of Physics, University of California, Berkeley, CA 94720, USA}
\author{Ken Van Tilburg} 
\email{kenvt@stanford.edu}
\affiliation{Stanford Institute for Theoretical Physics, Stanford University, Stanford, California 94305, USA}

\date{\today}

\begin{abstract}
We show that gravitational wave detectors based on a type of atom interferometry are sensitive to ultralight scalar dark matter.
Such dark matter can cause temporal oscillations in fundamental constants with a frequency set by the dark matter mass, and amplitude determined by the local dark matter density.
The result is a modulation of atomic transition energies.  This signal is ideally suited to a type of gravitational wave detector that compares two spatially separated atom interferometers referenced by a common laser.
Such a detector can improve on current searches for electron-mass or electric-charge modulus dark matter by up to 10 orders of magnitude in coupling, in a frequency band complementary to that of other proposals. 
It demonstrates that this class of atomic sensors is qualitatively different from other gravitational wave detectors, including those based on laser interferometry.  By using atomic-clock-like interferometers, laser noise is mitigated with only a single baseline.  These atomic sensors can thus detect scalar signals in addition to tensor signals.
\end{abstract}
\maketitle


\textit{Introduction.---}
The search for dark matter (DM) is one of the most important goals in particle physics.  There are now many experiments designed for the direct detection of DM.  Almost all of these search for heavier DM, with mass well above an eV, using energy deposition from DM particles scattering in the detector.  Traditional particle detection techniques have energy thresholds which make it challenging to look for lighter DM.  However, there is a vast range of DM parameter space with mass far below the level detectable in these experiments.  New types of technology are required to search for ultralight DM.  

The QCD axion is perhaps the best known example of light DM, but there are many other motivated possibilities such as light moduli~\cite{Dimopoulos:1996kp,ArkaniHamed:1999dz,Burgess:2010sy,Cicoli:2011yy}, dilatons~\cite{Damour:1994zq,Taylor:1988nw}, Higgs portal DM~\cite{Piazza:2010ye}, and the relaxion~\cite{Graham:2015cka} among many others.  We focus on DM with scalar couplings to matter which causes time variation of fundamental constants such as the electron mass \cite{Arvanitaki:2014faa}.  This type of DM can be searched for using atomic clocks \cite{Arvanitaki:2014faa, VanTilburg:2015oza, Hees:2016gop}, resonant-mass detectors~\cite{Arvanitaki:2015iga}, and accelerometers~\cite{Graham:2015ifn}.

In this Letter, we demonstrate that a class of atomic sensors for gravitational waves \cite{Graham:2012sy} can be used for direct detection of scalar DM over many orders of magnitude in mass.
This type of atomic sensor is unique in that it has full sensitivity with a single baseline because it relies on atom interferometers designed to be similar to atomic clocks.   As a single-baseline detector it does not rely on the tensor nature of the gravitational wave (GW).  This makes it ideal for searching for scalar DM as well.

\textit{Model.---}
A scalar DM particle will naturally couple to the Standard Model particles, and hence potentially be observable, through a relatively small number of couplings in the effective field theory.  
In this Letter, we consider a representative set of its couplings, described by the Lagrangian
\begin{align}
\mathcal{L} = &+ \half \partial_\mu \phi \partial^\mu \phi  - \half m_\phi^2 \phi^2 \label{eq:kinetic}\\
& - \sqrt{4\pi G_N} \phi \left[ d_{m_e} m_e \bar{e} e - \frac{d_e}{4} F_{\mu \nu} F^{\mu \nu} \right], \label{eq:couplings}
\end{align}
where we parametrized the leading interaction with electrons and photons relative to gravity as in Refs.~\cite{Damour:2010rm,Damour:2010rp}: $G_N$ is Newton's constant, so $d_{m_e} = d_e = 1$ would be the couplings of a scalar graviton. We employ units in which $\hbar = c = 1$. The couplings in Eq.~\ref{eq:couplings} could originate from a Higgs portal coupling of the form $\mathcal{L} \supset b \phi |H|^2$, which is one ultraviolet completion into a renormalizable model with a particularly low cutoff~\cite{Piazza:2010ye}. Scalar fields with quadratic couplings to matter \cite{PhysRevD.77.043524,PhysRevLett.114.161301}, e.g.~$\mathcal{L} \supset \phi^2 |H|^2$, give rise to analogous signatures as the linear couplings, but have drastically more fine-tuned masses for the same physical effect, so we shall not consider them further.

Bosonic DM much lighter than 1~eV is a highly classical state because of high occupation numbers, and can be approximated by a nonrelativistic plane wave solution to Eq.~\ref{eq:kinetic}:
\begin{align}
\phi\left(t,\mathbf{x}\right) = \phi_0 \cos\left[m_\phi (t -  \mathbf{v}\cdot\mathbf{x}) + \beta\right]+\mathcal{O}\left(|\mathbf{v}|^2\right), \label{eq:solution}
\end{align}
with amplitude $\phi_0 \simeq \sqrt{2\rho_\text{DM}}/{m_\phi}$ determined by the local DM energy density $\rho_\text{DM} \approx 0.3 \GeV / \text{cm}^{3}$. The local description of Eq.~\ref{eq:solution} should be thought of as an incoherent superposition of waves (hence $\phi_0 \propto \sqrt{\rho_\text{DM}}$) that nevertheless has a long phase coherence time of approximately $2 \pi / m_\phi v_\text{vir}^2$ where $v_\text{vir} \sim 10^{-3}$ is the Galactic virial velocity. The coherence arises from the nonrelativistic nature of DM: the angular frequency of the wave is mostly set by the rest-mass energy $m_\phi$. It receives small kinetic energy corrections of $\mathcal{O}(m_\phi v_\text{vir}^2)$, which do have a large spread: $\langle|\mathbf{v}|\rangle \sim v_\text{vir}$ and $\langle (\mathbf{v} - \langle \mathbf{v} \rangle)^2 \rangle \sim v_\text{vir}^2$.

The scalar field DM oscillations of Eq.~\ref{eq:solution} combined with the couplings to matter of Eq.~\ref{eq:couplings} cause fundamental ``constants'' such as the electron mass and the fine-structure constant to oscillate in time:
\begin{align}
m_e(t,\mathbf{x}) &= m_e \left[1 + d_{m_e} \sqrt{4\pi G_N} \phi(t,\mathbf{x})\right]\\
\alpha(t,\mathbf{x}) &= \alpha \left[1 + d_{e} \sqrt{4\pi G_N} \phi(t,\mathbf{x})\right].
\end{align}
Temporal variation of $m_e$ and $\alpha$ gives rise to oscillations in energy and length scales in atoms, phenomena which were respectively exploited by DM search proposals using atomic clock pairs~\cite{Arvanitaki:2014faa} and resonant-mass detectors~\cite{Arvanitaki:2015iga}. Spatial variation leads to oscillating, chemistry-dependent forces, which can be looked for with accelerometers~\cite{Graham:2015ifn} (see also~Ref.~\cite{Arvanitaki:2014faa} for a tidal-force effect). 

We will show that DM-induced temporal variation of atomic transition frequencies can be searched for with a \emph{single} atomic species by exploiting the time-domain response of a differential atomic interferometer to a scalar DM wave. The search strategy outlined below has discovery reach for scalar DM couplings that is potentially orders of magnitude better than existing constraints and proposals in its frequency band, and is complementary to the low-frequency, broadband strategies of Refs.~\cite{Arvanitaki:2014faa}~and~\cite{Graham:2015ifn}, and the high-frequency, resonant searches of Ref.~\cite{Arvanitaki:2015iga}.

An electronic transition energy $\omega_\text{A}$ depends on the values of $m_e$ and $\alpha$, and so will oscillate itself in the presence of a scalar DM wave:
\begin{align}
\omega_\text{A}(t) &\simeq \omega_\text{A} + \Delta \omega_\text{A} \cos(m_\phi t);\label{eq:omegaA}\\
\Delta \omega_\text{A} &\equiv \omega_\text{A} \sqrt{4\pi G_N} \phi_0 \left(d_{m_e} + \xi d_e\right).\label{eq:DeltaomegaA}
\end{align}
Above, we have neglected the $v_\text{vir}$-suppressed spatial variation and the $v_\text{vir}^2$-suppressed temporal incoherence, which we will restore in our final results. We assumed a linear dependence of $\omega_\text{A}$ on $m_e$, valid to a high degree for all (nonhyperfine) electronic transitions. For the $\rm 5s^2 \, {}^1S_0 \leftrightarrow \rm 5s5p \, {}^3P_0$ transition in Sr~I, which we will take as a case study throughout, one has $\xi \approx 2.06$~\cite{PhysRevA.70.014102}.

\begin{figure}[t]
\includegraphics[width = 0.48 \textwidth]{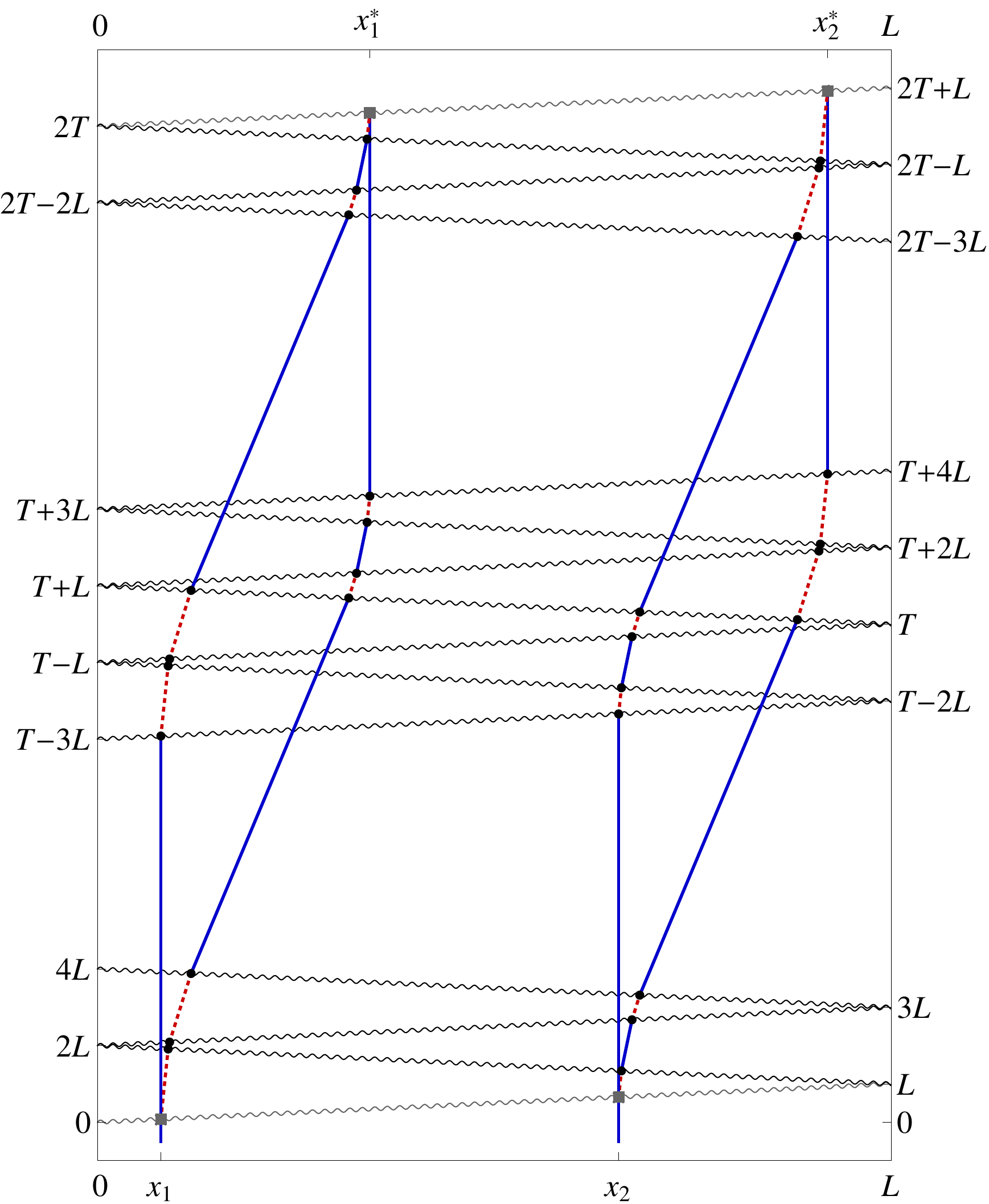}
\caption{Spacetime diagram of the light-pulse sequence on two atom interferometers, illustrated for $n=4$ (i.e.~maximum 4 photon momenta transferred). A $\pi/2$ laser pulse (gray, wavy) splits the wavefunction of atoms both at $x_1$ and at $x_2$ into the ground state $|g\rangle$ (blue, solid) and the excited state $|e\rangle$ (red, dashed) with equal probabilities. Subsequent $\pi$ laser pulses (black, wavy) exchange $|g\rangle \leftrightarrow |e\rangle$, and typically only interact with one branch of each wavefunction due to Doppler shifts. A final $\pi/2$ pulse interferes the wavefunction of both interferometers. Interaction points are indicated by gray squares (black dots) for $\pi/2$ ($\pi$) pulses. For clarity, atomic separations are exaggerated; realistically, $x_1 \sim x_1^* \sim L-x_2 \sim L-x_2^* \ll L$.
}\label{fig:setup}
\end{figure}

\textit{Physical effect.---}
The light-pulse atom interferometry scheme of \cite{Graham:2012sy}, depicted in Fig.~\ref{fig:setup}, is like a \emph{differential} atomic clock, where \emph{one} laser is referenced to \emph{two} spatially separated atomic ensembles. In absence of new physics and reducible backgrounds, the phase response in the atomic ensembles is identical, but both suffer from laser noise imprinted onto the atoms, especially at frequencies below 1~Hz. However, the differential atomic phase response is insensitive to laser noise when the atoms move along free-fall geodesics when not manipulated by the laser. This differential phase response can serve as a low-background channel to look for new physics. A GW would modulate the light travel time of the laser pulses between the atomic ensembles, leading to a differential phase accumulation over the interferometer sequence~\cite{Graham:2012sy}.

Atomic sensors for GW detection are also intrinsically sensitive to modulus DM waves, without change to the experimental configuration. Given that all phases in the sequence of Fig.~\ref{fig:setup} cancel in absence of new physics, we keep track only of DM-induced phase accumulation $\Phi$ of the excited atomic state relative to that of the ground state. Between times $t_0$ and $t_1$, this amounts to:
\begin{align}
\Phi_{t_0}^{t_1} \equiv \int_{t_0}^{t_1} dt\, \Delta \omega_\text{A} \cos(m_\phi t + \beta ) .
\end{align}
The signal channel constitutes of the signal phase $\Phi_\text{s}$, the phase difference of an atom interferometer located at $x_1 \simeq 0$ subtracted by that of one at $x_2 \simeq L$. For the setup in Fig.~\ref{fig:setup}, this is approximately equal to
\begin{align}
\Phi_\text{s} \simeq \Phi_{T-(n-1)L}^{T+L} - \Phi_0^{nL}
- \Phi_{2T-(n-1)L}^{2T+L} + \Phi_{T}^{T+nL},\label{eq:signalphase}
\end{align}
where $T$ is half the time between the two $\pi/2$ beamsplitter pulses, $L \simeq x_2 - x_1$ is the light travel time between the two laser sources, and $n$ is the number of large-momentum-transfer (LMT) photon kicks each atom receives. In the limits of $m_\phi \to 0$ and $m_\phi \to \infty$, $\Phi_\text{s}$ asymptotes to zero. However, a nontrivial signal phase response does occur when the period of the DM wave matches the total duration of the interferometric sequence, namely $2\pi / m_\phi \sim 2 T$. (By construction, $T > n L$, and $T \gg L$ for the setups under consideration.) For example, in the optimally-matched case with a DM phase $\beta = 0$ at the start of the interferometric sequence, all of the terms in Eq.~\ref{eq:signalphase} are negative, because terms 2 and 3 (1 and 4) are generated during positive (negative) anti-nodes of the DM wave, yielding a signal phase shift of order $\Phi_s \sim - 4 \Delta \omega_\text{A}(n L)$. The signal amplitude of Eq.~\ref{eq:signalphase}, $\overline{\Phi}_\text{s} \equiv ( 2 \int_0^{2\pi} d\beta \,  \Phi_\text{s}^2 / 2\pi )^{1/2}$, for general $m_\phi$ is
\begin{align}
\overline{\Phi}_\text{s} & = 8 \frac{\Delta \omega_A}{m_\phi} \label{eq:signalphaseavg}\\
& \times \left| \sin\left[\frac{m_\phi n L}{2}\right] \sin\left[\frac{m_\phi (T-(n-1)L}{2} \right] \sin\left[\frac{m_\phi T}{2} \right]\right|.\nonumber
\end{align}
Since $\Delta \omega_\text{A} \propto 1/m_\phi$ at fixed DM energy density (see below Eq.~\ref{eq:solution}), we deduce that the effect decouples $\propto m_\phi$ for $m_\phi \to 0$, and $\propto 1/m_\phi^2$ for $m_\phi \to \infty$. 

The experiment under consideration can be thought of as a comparison of two atomic clocks.  Here the clocks are spatially separated, which is what makes it a GW detector.  This also creates a difference in the effect of the scalar DM on the two clocks, allowing a differential measurement to cancel laser noise but not the DM signal.
This observable effect differs from that of other proposed experiments searching for scalar DM using atomic clock-based technology.
For example, our proposal shows that the scalar DM effect can be detected using a \emph{single species} of atoms, in contrast to the methods of Ref.~\cite{Arvanitaki:2014faa}. This is because the differential setup of Ref.~\cite{Graham:2012sy} allows us to compare the response of two otherwise identical atomic clocks at different points in time, which is kept by the phase evolution of the DM wave. The advantages of GW sensors of the type described in Ref.~\cite{Graham:2012sy} will thus improve DM searches by many orders of magnitude over a wide range of masses. 

Additionally, our proposal differs from the recent Ref.~\cite{Geraci:2016fva}, which also proposed atom interferometer GW detectors for scalar DM detection.  We consider the direct effect of the scalar DM on the \emph{internal} state of the atomic ensemble while Ref.~\cite{Geraci:2016fva} mainly relies on the DM effect on the Earth's gravitational field. Our proposal achieves best discovery potential in the most sensitive frequency band of GW detectors, while Ref.~\cite{Geraci:2016fva} is sensitive only to lower-frequency signals.

\begin{figure}[t]
\includegraphics[trim={0 2.0cm 0 0},clip,width = 0.48 \textwidth]{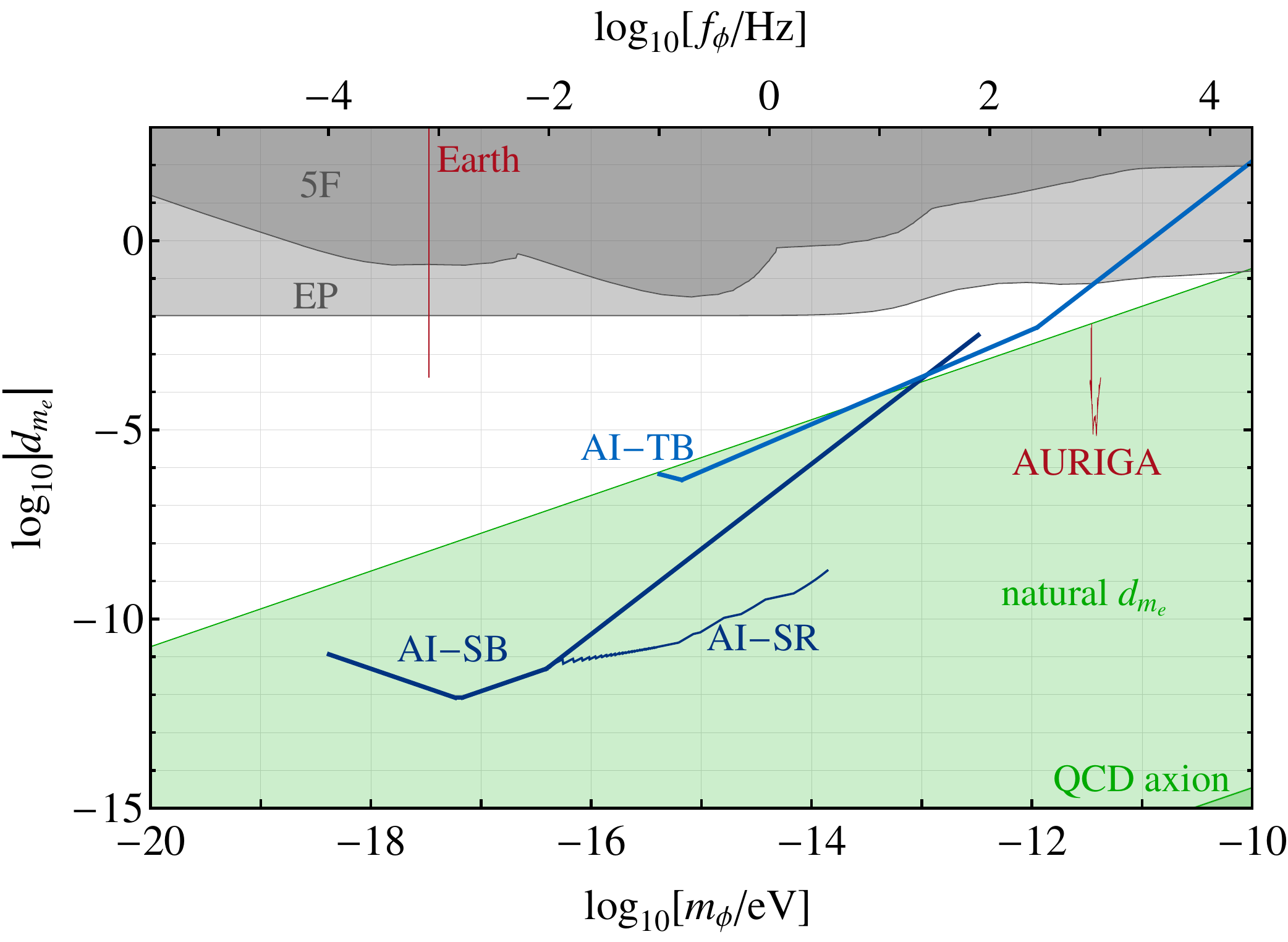}
\includegraphics[trim={0 0 0 2.0cm},clip,width = 0.48 \textwidth]{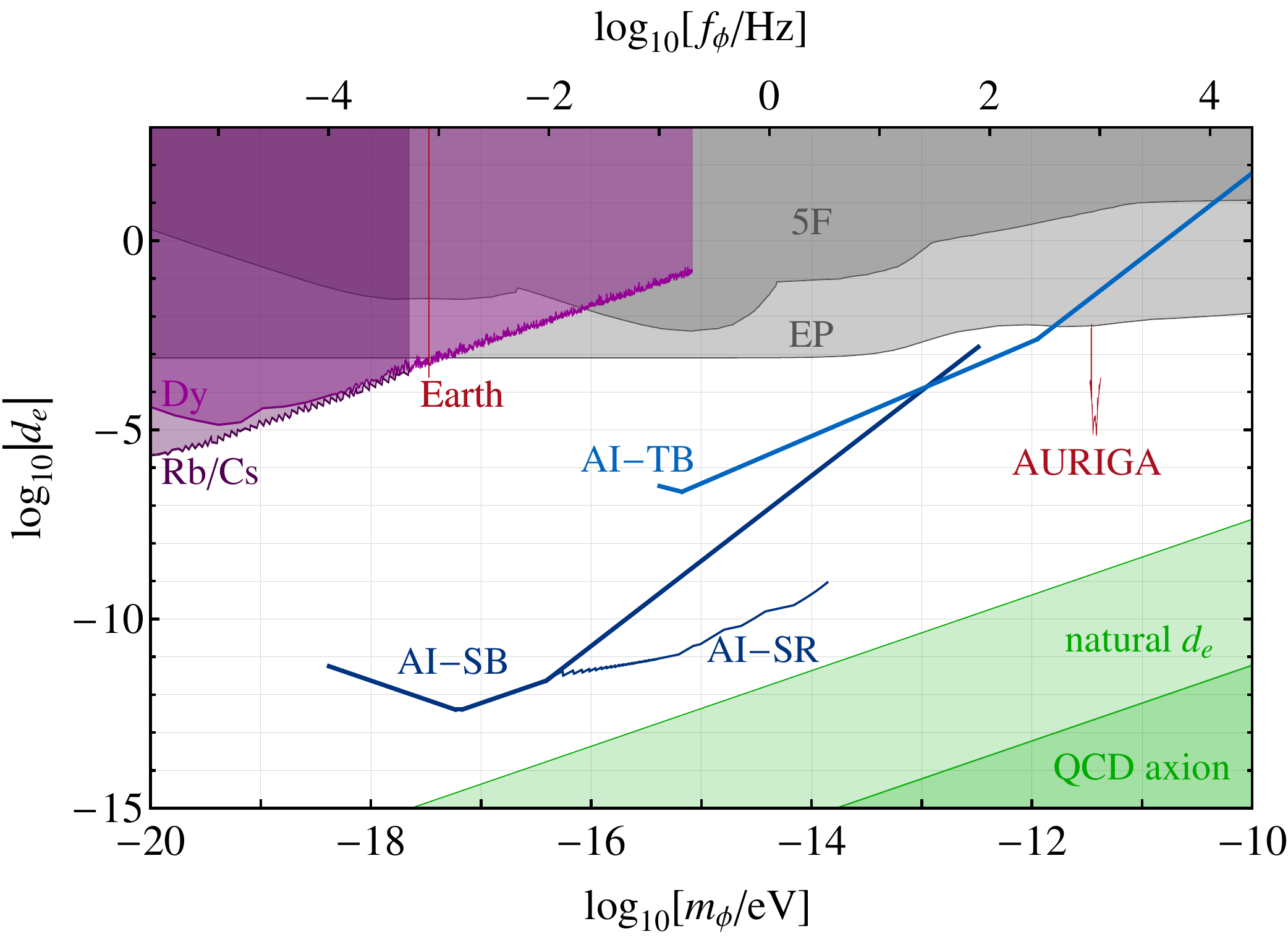}
\caption{Parameter space for the coupling $d_{m_e}$ to electrons (top panel) and $d_e$ to photons (bottom panel), as a function of dark matter mass $m_\phi$. Blue curves depict the $\text{SNR}=1$ sensitivity envelopes of the proposed atomic sensors: a terrestrial experiment operated in broadband mode (``AI-TB''), long-baseline, broadband, space-based antenna (``AI-SB''), and a shorter, resonant satellite antenna (``AI-SR'').  Also depicted are 95\%-CL constraints from searches for new Yukawa forces that violate/conserve the equivalence principle (``EP/5F'', gray regions), atomic spectroscopy data in Dy and in Rb/Cs (light and dark purple regions), and seismic data on the fundamental breathing mode of Earth (red). The potential reach for an analysis on existing AURIGA data, representative of the sensitivity of resonant-mass detectors, is also shown in red. Green regions show natural parameter space for a 10-TeV cutoff, and allowed parameter space for the QCD axion.
}\label{fig:coupling}
\end{figure}

\textit{Experimental sensitivity.---}
The scheme proposed in Ref.~\cite{Graham:2012sy} can be realized in a ground-based interferometer as well as in a space-based satellite antenna. A terrestrial experiment could be operated in a vertical shaft of length $L = 10^3$~m with 10-meter interferometers at the top and bottom, allowing free-fall times of $T = 1.4~\text{s}$. 
We restrict to a maximum number of $N_\text{max} = 10^3$ laser pulses in order to retain atom number, which in turn limits the number of LMT kicks to $n = 250$.
We assume shot-noise-limited sensitivity above $f= 10^{-1}~\text{Hz}$ with a noise spectral density $\sqrt{S_\Phi} \approx 10^{-5}/\text{Hz}^{1/2}$, made possible with an atomic flux of $10^{10}/\text{s}$, or with fewer atoms and significant squeezing~\cite{hosten2016measurement}.

A space-based satellite experiment can exhibit a much longer baseline length $L$ and interrogation time $T$, because the laser platforms can move on free-fall geodesics along with the atoms. A GW antenna design using atom interferometry near satellites connected with heterodyne laser links~\cite{Hogan:2015xla} has a proposed configuration with $L = 6 \times 10^8~\text{m}$, $T = 160~\text{s}$, and $n = 12$. The baseline length and interrogation time are limited by laser diffraction, and atomic loss due to scattering with background gas and light, respectively. For this setup, we assume a shot-noise-limited sensitivity of $\sqrt{S_\Phi} \approx 10^{-4}/\text{Hz}^{1/2}$.

Given a DM signal bandwidth of $\Delta f_\phi \simeq m_\phi v_\text{vir}^2 / 2\pi$, differential atomic phase oscillations with square amplitude as small as $\delta \Phi_\text{s}^2 = S_\Phi t_\text{int}^{-1} \max\lbrace 1, t_\text{int} \Delta f_\phi \rbrace^{1/2}$ may be detected at unit signal-to-noise ratio ($\text{SNR} = 1$) after an integration time $t_\text{int} = 10^8~\text{s}$. With the parameters for $L$, $T$, $n$, and the atomic transition (throughout assumed to be the ${}^{1}\rm{S}_0 \leftrightarrow {}^3\rm{P}_0$ transition in Sr~I), the discovery reach for DM couplings can then be computed as a function $m_\phi$ with aid of Eqs.~\ref{eq:DeltaomegaA}~and~\ref{eq:signalphaseavg}.

\begin{figure}[t]
\includegraphics[trim={0 0 0 0},clip,width = 0.48 \textwidth]{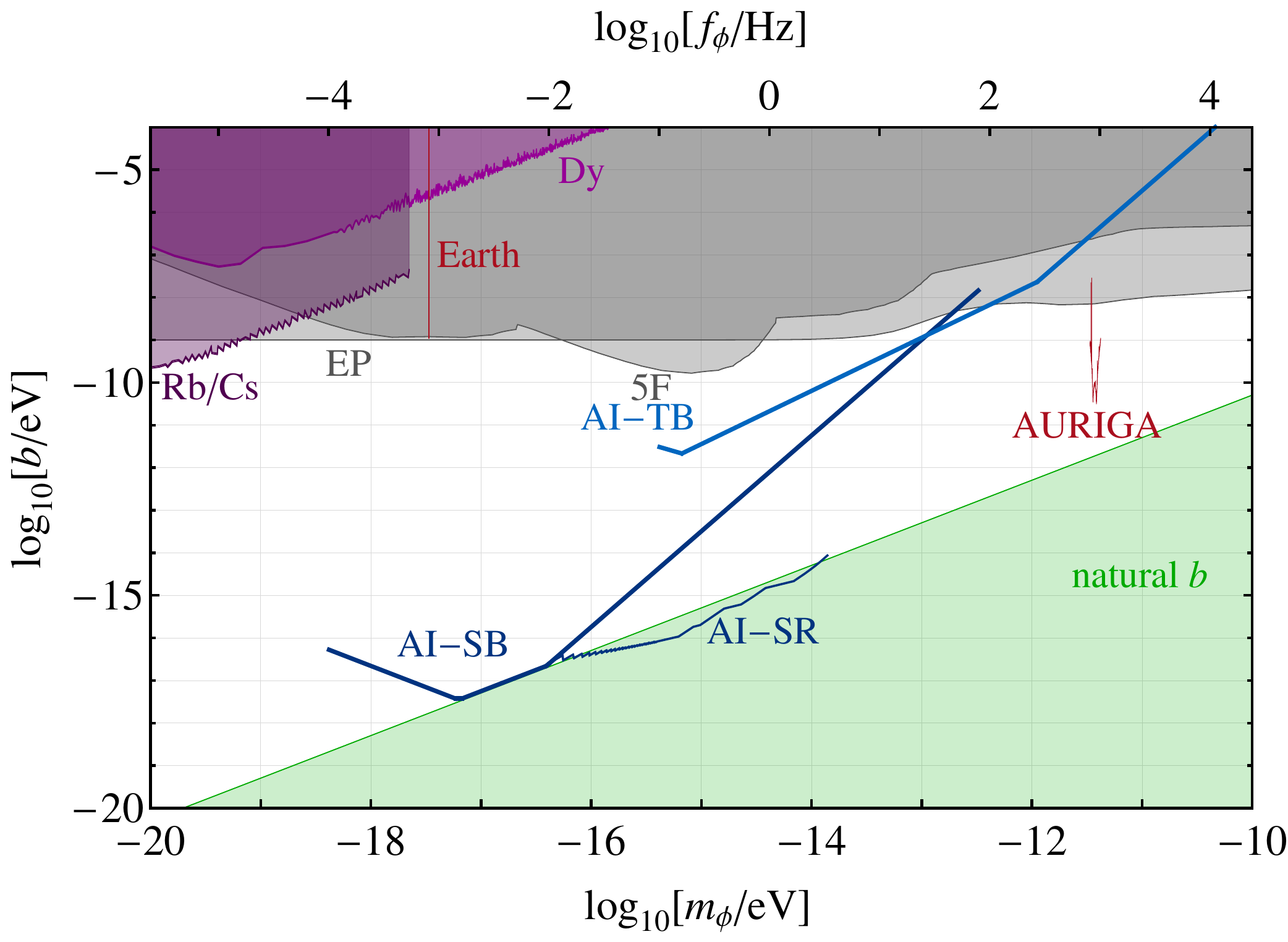}
\caption{Parameter space for the Higgs portal coupling $b$ as a function of dark matter mass $m_\phi$. Curves and regions are as in Fig.~\ref{fig:coupling}. Here the green region highlights couplings $b$ for which the lightest physical mass eigenvalue $m_\phi$ of the scalar potential is natural at the classical level, as described in the text. Loop-level quantum corrections to the mass are subdominant, so the natural region is independent of the UV cutoff.
}\label{fig:couplingb}
\end{figure}

Atomic sensors provide extraordinary discovery reach, with a potential to improve on existing constraints and other proposals by many orders of magnitude over a wide frequency band. In Fig.~\ref{fig:coupling}, we plot the sensitivity to the electron coupling $d_{m_e}$ (top panel) and photon coupling $d_e$ (bottom panel) for both the terrestrial (``AI-TB'', light blue) and space-based proposals (``AI-SB'', dark blue). Analogous curves for the Higgs portal coupling are plotted in Fig.~\ref{fig:couplingb}. For clarity, we used the approximation $|\sin(x)| \sim \min\lbrace x,1/\sqrt{2} \rbrace$ for the power-averaged envelope of Eq.~\ref{eq:signalphaseavg}. 
We note that a DM signal with frequency above the repetition frequency of the interferometer sequence (typically about 1~Hz) will be aliased to lower frequencies, and can still be detected with the same phase sensitivity over the parameter space of interest. We show the reach of the space-based proposal up to frequencies where the DM wave becomes spatially incoherent (when $m_\phi v_\text{vir} L \gtrsim 1$) and down to frequencies where gravity gradients are deemed to become more important than shot noise, at $f \lesssim 10^{-4}~\text{Hz}$. For the ground-based proposal, gravity gradients can likely be kept subdominant for $f \gtrsim 10^{-1}~\text{Hz}$~\cite{Dimopoulos:2007cj,Dimopoulos:2008sv}.

In Figs.~\ref{fig:coupling} and \ref{fig:couplingb}, we also show 95\%-CL gray exclusion regions from equivalence-principle tests~\cite{Schlamminger:2007ht,Wagner:2012ui} and searches for a Yukawa-type deviation from the gravitational force~\cite{Adelberger:2003zx}, which are both independent of DM abundance. Atomic spectroscopy limits at 95\% CL on oscillations of relative transition energies in isotopes of Dy~\cite{VanTilburg:2015oza}, and in Rb and Cs~\cite{Hees:2016gop} are also shown. In red, we plot results of atomic length scale oscillation effects~\cite{Arvanitaki:2015iga}: a limit derived from terrestrial seismic data~\cite{JGR:JGR4118}, and a prospective reach utilizing existing data on the resonant-mass detector AURIGA~\cite{auriga1}. 

Green regions in Fig.~\ref{fig:coupling} indicate natural parameter space--where loop-level quantum corrections to the scalar mass are less than the physical mass $m_\phi$ for an ultraviolet cutoff of 10~TeV--as well as allowed parameter space for the QCD axion.  In Fig.~\ref{fig:couplingb}, the green region highlights natural Higgs portal couplings---regardless of UV cutoff. Elsewhere, the coupling $b$ is tuned against the bare mass of $\phi$ at the classical level (loop corrections are subdominant), such that the lightest mass eigenstate in the $\phi$--$H$ scalar potential has a physical mass $m_\phi \lesssim b / \sqrt{2 \lambda}$. This mass eigenstate is rotated slightly in the Higgs direction, by an angle $b / \sqrt{2 \lambda m_h^2}$ with $m_h$ the heavy mass eigenvalue at 125~GeV and the Higgs quartic normalized as $\mathcal{L} \supset \lambda |H|^4$, leading to e.g.~the correspondence $ d_{m_e} \sqrt{4\pi G_N} \simeq b / m_h^2$~\cite{Piazza:2010ye,Graham:2015ifn}.

The atomic sensors of the type described in Ref.~\cite{Graham:2012sy} are broadband sensors, but they can also be operated in resonant mode. By interweaving many diamond-shaped atomic paths of the type in Fig.~\ref{fig:setup} in a fixed interferometer duration, the detectors become resonantly sensitive to higher-frequency signals~\cite{Graham:2016plp}. With a $Q_d$ number of diamonds in a total sequence duration $\tilde{T}$, maximum sensitivity is achieved at $f_d \sim Q_d/\tilde{T}$ (and integer multiples thereof) in a frequency band $\Delta f_d \sim f_d/Q_d$. Constrained on keeping the total time below $T_\text{max} = 300$~s and the total number of laser pulses below $N_\text{max} = 10^3$, a strain spectral noise density of $\sqrt{S_\text{h}} \sim 10^{-22}/\text{Hz}^{1/2}$ may be attained with a baseline of $L = 4.4\times10^7~\text{m}$ and phase noise $\sqrt{S_\Phi} \approx 10^{-5}/\text{Hz}^{1/2}$ in a frequency range between $f_\text{min} = 0.01~\text{Hz}$ and $f_\text{max} =  4~\text{Hz}$~\cite{Graham:2016plp}.

Sweeping through this band with a constant average fractional frequency scanning speed of $\sigma \equiv \Delta f / f \Delta t$ by changing $Q_d$ and $\tilde{T}$ would take an integration time of $t_\text{int} = \sigma^{-1} \ln(f_\text{max}/f_\text{min})$, again taken to be $10^8~\text{s}$. This yields a reach for e.g.~$d_{m_e}$ of approximately $\sqrt{S_\text{h}\sigma f/f_\text{min}}/2\sqrt{4\pi G_N} \phi_0$, which is plotted as the thin blue curve (``AI-SR'') in the top panel Fig.~\ref{fig:coupling}, with analogous results for $d_e$ and $b$ in Figs.~\ref{fig:coupling}~and~\ref{fig:couplingb}. The resonant mode would allow precise dissection of a positive narrowband signal. It can also be realized in the above terrestrial detector, though with fewer sensitivity benefits of a scanning search relative to broadband operation. 

Atomic GW detectors also have significant sensitivity to topological defects of fields with scalar couplings~\cite{Derevianko:2013oaa}, through two separate physical effects, both yielding transient signals. The first arises when one atom interferometer is inside a defect, and thus has its atomic transition energies shifted relative to those of the other interferometer. The second effect results from the field gradient at the defect edges, producing differential forces~\cite{Arvanitaki:2014faa,Graham:2015ifn} and thus apparent strains on the interferometers. The second signature is present in any GW detector with free-falling test masses, including aLIGO~\cite{0264-9381-32-7-074001}. Our preliminary estimates show that GW detectors have a potential sensitivity much beyond that of atomic clock experiments~\cite{Derevianko:2013oaa, 2016arXiv160505763W}, warranting a detailed analysis elsewhere.

\textit{Conclusion.---}
Most GW detection techniques based on interferometry rely on the tensor nature of the GW, and have significantly reduced sensitivity to scalar signals, including those from scalar DM. Laser interferometers such as aLIGO~\cite{0264-9381-32-7-074001} or the proposed eLISA~\cite{LISA}, and atom interferometers such as AGIS~\cite{Dimopoulos:2007cj, Dimopoulos:2008sv, Hogan:2010fz} compare the signal in multiple directions in order to cancel out laser noise, which would otherwise severely limit the sensitivity. Laser frequency noise is similar to the scalar DM effect, since it acts on both baselines in the same way. By looking for a differential response on multiple equal-length baselines, these GW detectors drastically reduce any signal from scalar DM. 

However, the GW detector described in Ref.~\cite{Graham:2012sy} uses a differential measurement of two atom interferometers designed to be similar to atomic clocks.  The use of atomic clock-like interferometers allows the removal of laser noise along a single baseline, unlike in many other GW detectors.  It is thus ideally suited for a scalar DM search. In this Letter, we identified a new signature of scalar dark matter in an atomic sensor of this type, and outlined search strategies in a wide range of natural parameter space for well-motivated dark matter candidates.

\acknowledgments{
\textit{Acknowledgments.---}
We thank James Thompson and Jun Ye for helpful discussions. AA acknowledges the support of NSERC and the Stavros Niarchos Foundation. Research at Perimeter Institute is supported by the Government of Canada through Industry Canada and by the Province of Ontario through the Ministry of Economic Development \& Innovation. 
PWG acknowledges the support of NSF grant PHY-1316706, DOE Early Career Award DE-SC0012012, and the W.M.~Keck Foundation.  SR was supported in part by the NSF under grants PHY-1417295 and PHY-1507160 and the Simons Foundation Award 378243. This work was supported in part by the Heising-Simons Foundation grants 2015-037 and 2015-038.
}


\bibliographystyle{apsrev4-1-etal}
\bibliography{aidm}

\begin{thebibliography}{35}%
\makeatletter
\providecommand \@ifxundefined [1]{%
 \@ifx{#1\undefined}
}%
\providecommand \@ifnum [1]{%
 \ifnum #1\expandafter \@firstoftwo
 \else \expandafter \@secondoftwo
 \fi
}%
\providecommand \@ifx [1]{%
 \ifx #1\expandafter \@firstoftwo
 \else \expandafter \@secondoftwo
 \fi
}%
\providecommand \natexlab [1]{#1}%
\providecommand \enquote  [1]{``#1''}%
\providecommand \bibnamefont  [1]{#1}%
\providecommand \bibfnamefont [1]{#1}%
\providecommand \citenamefont [1]{#1}%
\providecommand \href@noop [0]{\@secondoftwo}%
\providecommand \href [0]{\begingroup \@sanitize@url \@href}%
\providecommand \@href[1]{\@@startlink{#1}\@@href}%
\providecommand \@@href[1]{\endgroup#1\@@endlink}%
\providecommand \@sanitize@url [0]{\catcode `\\12\catcode `\$12\catcode
  `\&12\catcode `\#12\catcode `\^12\catcode `\_12\catcode `\%12\relax}%
\providecommand \@@startlink[1]{}%
\providecommand \@@endlink[0]{}%
\providecommand \url  [0]{\begingroup\@sanitize@url \@url }%
\providecommand \@url [1]{\endgroup\@href {#1}{\urlprefix }}%
\providecommand \urlprefix  [0]{URL }%
\providecommand \Eprint [0]{\href }%
\providecommand \doibase [0]{http://dx.doi.org/}%
\providecommand \selectlanguage [0]{\@gobble}%
\providecommand \bibinfo  [0]{\@secondoftwo}%
\providecommand \bibfield  [0]{\@secondoftwo}%
\providecommand \translation [1]{[#1]}%
\providecommand \BibitemOpen [0]{}%
\providecommand \bibitemStop [0]{}%
\providecommand \bibitemNoStop [0]{.\EOS\space}%
\providecommand \EOS [0]{\spacefactor3000\relax}%
\providecommand \BibitemShut  [1]{\csname bibitem#1\endcsname}%
\let\auto@bib@innerbib\@empty
\bibitem [{\citenamefont {Dimopoulos}\ and\ \citenamefont
  {Giudice}(1996)}]{Dimopoulos:1996kp}%
  \BibitemOpen
  \bibfield  {author} {\bibinfo {author} {\bibfnamefont {S.}~\bibnamefont
  {Dimopoulos}}\ and\ \bibinfo {author} {\bibfnamefont {G.~F.}\ \bibnamefont
  {Giudice}},\ }\bibfield  {booktitle} {\emph {\bibinfo {booktitle} {{ITP
  Workshop on SUSY Phenomena and SUSY GUTS Santa Barbara, California, December
  7-9, 1995}}},\ }\href {\doibase 10.1016/0370-2693(96)00390-5} {\bibfield
  {journal} {\bibinfo  {journal} {Phys. Lett.}\ }\textbf {\bibinfo {volume}
  {B379}},\ \bibinfo {pages} {105} (\bibinfo {year} {1996})},\ \Eprint
  {http://arxiv.org/abs/hep-ph/9602350} {arXiv:hep-ph/9602350 [hep-ph]}
  \BibitemShut {NoStop}%
\bibitem [{\citenamefont {Arkani-Hamed}\ \emph {et~al.}(2000)\citenamefont
  {Arkani-Hamed}, \citenamefont {Hall}, \citenamefont {Smith},\ and\
  \citenamefont {Weiner}}]{ArkaniHamed:1999dz}%
  \BibitemOpen
  \bibfield  {author} {\bibinfo {author} {\bibfnamefont {N.}~\bibnamefont
  {Arkani-Hamed}}, \bibinfo {author} {\bibfnamefont {L.~J.}\ \bibnamefont
  {Hall}}, \bibinfo {author} {\bibfnamefont {D.}~\bibnamefont {Smith}}, \ and\
  \bibinfo {author} {\bibfnamefont {N.}~\bibnamefont {Weiner}},\ }\href
  {\doibase 10.1103/PhysRevD.62.105002} {\bibfield  {journal} {\bibinfo
  {journal} {Phys. Rev.}\ }\textbf {\bibinfo {volume} {D62}},\ \bibinfo {pages}
  {105002} (\bibinfo {year} {2000})},\ \Eprint
  {http://arxiv.org/abs/hep-ph/9912453} {arXiv:hep-ph/9912453 [hep-ph]}
  \BibitemShut {NoStop}%
\bibitem [{\citenamefont {Burgess}\ \emph {et~al.}(2011)\citenamefont
  {Burgess}, \citenamefont {Maharana},\ and\ \citenamefont
  {Quevedo}}]{Burgess:2010sy}%
  \BibitemOpen
  \bibfield  {author} {\bibinfo {author} {\bibfnamefont {C.~P.}\ \bibnamefont
  {Burgess}}, \bibinfo {author} {\bibfnamefont {A.}~\bibnamefont {Maharana}}, \
  and\ \bibinfo {author} {\bibfnamefont {F.}~\bibnamefont {Quevedo}},\ }\href
  {\doibase 10.1007/JHEP05(2011)010} {\bibfield  {journal} {\bibinfo  {journal}
  {JHEP}\ }\textbf {\bibinfo {volume} {05}},\ \bibinfo {pages} {010} (\bibinfo
  {year} {2011})},\ \Eprint {http://arxiv.org/abs/1005.1199} {arXiv:1005.1199
  [hep-th]} \BibitemShut {NoStop}%
\bibitem [{\citenamefont {Cicoli}\ \emph {et~al.}(2011)\citenamefont {Cicoli},
  \citenamefont {Burgess},\ and\ \citenamefont {Quevedo}}]{Cicoli:2011yy}%
  \BibitemOpen
  \bibfield  {author} {\bibinfo {author} {\bibfnamefont {M.}~\bibnamefont
  {Cicoli}}, \bibinfo {author} {\bibfnamefont {C.~P.}\ \bibnamefont {Burgess}},
  \ and\ \bibinfo {author} {\bibfnamefont {F.}~\bibnamefont {Quevedo}},\ }\href
  {\doibase 10.1007/JHEP10(2011)119} {\bibfield  {journal} {\bibinfo  {journal}
  {JHEP}\ }\textbf {\bibinfo {volume} {10}},\ \bibinfo {pages} {119} (\bibinfo
  {year} {2011})},\ \Eprint {http://arxiv.org/abs/1105.2107} {arXiv:1105.2107
  [hep-th]} \BibitemShut {NoStop}%
\bibitem [{\citenamefont {Damour}\ and\ \citenamefont
  {Polyakov}(1994)}]{Damour:1994zq}%
  \BibitemOpen
  \bibfield  {author} {\bibinfo {author} {\bibfnamefont {T.}~\bibnamefont
  {Damour}}\ and\ \bibinfo {author} {\bibfnamefont {A.~M.}\ \bibnamefont
  {Polyakov}},\ }\href {\doibase 10.1016/0550-3213(94)90143-0} {\bibfield
  {journal} {\bibinfo  {journal} {Nucl. Phys.}\ }\textbf {\bibinfo {volume}
  {B423}},\ \bibinfo {pages} {532} (\bibinfo {year} {1994})},\ \Eprint
  {http://arxiv.org/abs/hep-th/9401069} {arXiv:hep-th/9401069 [hep-th]}
  \BibitemShut {NoStop}%
\bibitem [{\citenamefont {Taylor}\ and\ \citenamefont
  {Veneziano}(1988)}]{Taylor:1988nw}%
  \BibitemOpen
  \bibfield  {author} {\bibinfo {author} {\bibfnamefont {T.}~\bibnamefont
  {Taylor}}\ and\ \bibinfo {author} {\bibfnamefont {G.}~\bibnamefont
  {Veneziano}},\ }\href {\doibase 10.1016/0370-2693(88)91290-7} {\bibfield
  {journal} {\bibinfo  {journal} {Phys.Lett.}\ }\textbf {\bibinfo {volume}
  {B213}},\ \bibinfo {pages} {450} (\bibinfo {year} {1988})}\BibitemShut
  {NoStop}%
\bibitem [{\citenamefont {Piazza}\ and\ \citenamefont
  {Pospelov}(2010)}]{Piazza:2010ye}%
  \BibitemOpen
  \bibfield  {author} {\bibinfo {author} {\bibfnamefont {F.}~\bibnamefont
  {Piazza}}\ and\ \bibinfo {author} {\bibfnamefont {M.}~\bibnamefont
  {Pospelov}},\ }\href {\doibase 10.1103/PhysRevD.82.043533} {\bibfield
  {journal} {\bibinfo  {journal} {Phys.Rev.}\ }\textbf {\bibinfo {volume}
  {D82}},\ \bibinfo {pages} {043533} (\bibinfo {year} {2010})},\ \Eprint
  {http://arxiv.org/abs/1003.2313} {arXiv:1003.2313 [hep-ph]} \BibitemShut
  {NoStop}%
\bibitem [{\citenamefont {Graham}\ \emph
  {et~al.}(2015{\natexlab{a}})\citenamefont {Graham}, \citenamefont {Kaplan},\
  and\ \citenamefont {Rajendran}}]{Graham:2015cka}%
  \BibitemOpen
  \bibfield  {author} {\bibinfo {author} {\bibfnamefont {P.~W.}\ \bibnamefont
  {Graham}}, \bibinfo {author} {\bibfnamefont {D.~E.}\ \bibnamefont {Kaplan}},
  \ and\ \bibinfo {author} {\bibfnamefont {S.}~\bibnamefont {Rajendran}},\
  }\href {\doibase 10.1103/PhysRevLett.115.221801} {\bibfield  {journal}
  {\bibinfo  {journal} {Phys. Rev. Lett.}\ }\textbf {\bibinfo {volume} {115}},\
  \bibinfo {pages} {221801} (\bibinfo {year} {2015}{\natexlab{a}})},\ \Eprint
  {http://arxiv.org/abs/1504.07551} {arXiv:1504.07551 [hep-ph]} \BibitemShut
  {NoStop}%
\bibitem [{\citenamefont {Arvanitaki}\ \emph {et~al.}(2015)\citenamefont
  {Arvanitaki}, \citenamefont {Huang},\ and\ \citenamefont
  {Van~Tilburg}}]{Arvanitaki:2014faa}%
  \BibitemOpen
  \bibfield  {author} {\bibinfo {author} {\bibfnamefont {A.}~\bibnamefont
  {Arvanitaki}}, \bibinfo {author} {\bibfnamefont {J.}~\bibnamefont {Huang}}, \
  and\ \bibinfo {author} {\bibfnamefont {K.}~\bibnamefont {Van~Tilburg}},\
  }\href {\doibase 10.1103/PhysRevD.91.015015} {\bibfield  {journal} {\bibinfo
  {journal} {Phys.Rev.}\ }\textbf {\bibinfo {volume} {D91}},\ \bibinfo {pages}
  {015015} (\bibinfo {year} {2015})},\ \Eprint {http://arxiv.org/abs/1405.2925}
  {arXiv:1405.2925 [hep-ph]} \BibitemShut {NoStop}%
\bibitem [{\citenamefont {Van~Tilburg}\ \emph {et~al.}(2015)\citenamefont
  {Van~Tilburg}, \citenamefont {Leefer}, \citenamefont {Bougas},\ and\
  \citenamefont {Budker}}]{VanTilburg:2015oza}%
  \BibitemOpen
  \bibfield  {author} {\bibinfo {author} {\bibfnamefont {K.}~\bibnamefont
  {Van~Tilburg}}, \bibinfo {author} {\bibfnamefont {N.}~\bibnamefont {Leefer}},
  \bibinfo {author} {\bibfnamefont {L.}~\bibnamefont {Bougas}}, \ and\ \bibinfo
  {author} {\bibfnamefont {D.}~\bibnamefont {Budker}},\ }\href {\doibase
  10.1103/PhysRevLett.115.011802} {\bibfield  {journal} {\bibinfo  {journal}
  {Phys. Rev. Lett.}\ }\textbf {\bibinfo {volume} {115}},\ \bibinfo {pages}
  {011802} (\bibinfo {year} {2015})},\ \Eprint
  {http://arxiv.org/abs/1503.06886} {arXiv:1503.06886 [physics.atom-ph]}
  \BibitemShut {NoStop}%
\bibitem [{\citenamefont {Hees}\ \emph {et~al.}(2016)\citenamefont {Hees},
  \citenamefont {Gena}, \citenamefont {Abgrall}, \citenamefont {Bize},\ and\
  \citenamefont {Wolf}}]{Hees:2016gop}%
  \BibitemOpen
  \bibfield  {author} {\bibinfo {author} {\bibfnamefont {A.}~\bibnamefont
  {Hees}}, \bibinfo {author} {\bibfnamefont {J.}~\bibnamefont {Gena}}, \bibinfo
  {author} {\bibfnamefont {M.}~\bibnamefont {Abgrall}}, \bibinfo {author}
  {\bibfnamefont {S.}~\bibnamefont {Bize}}, \ and\ \bibinfo {author}
  {\bibfnamefont {P.}~\bibnamefont {Wolf}},\ }\href@noop {} {\  (\bibinfo
  {year} {2016})},\ \Eprint {http://arxiv.org/abs/1604.08514} {arXiv:1604.08514
  [gr-qc]} \BibitemShut {NoStop}%
\bibitem [{\citenamefont {Arvanitaki}\ \emph {et~al.}(2016)\citenamefont
  {Arvanitaki}, \citenamefont {Dimopoulos},\ and\ \citenamefont
  {Van~Tilburg}}]{Arvanitaki:2015iga}%
  \BibitemOpen
  \bibfield  {author} {\bibinfo {author} {\bibfnamefont {A.}~\bibnamefont
  {Arvanitaki}}, \bibinfo {author} {\bibfnamefont {S.}~\bibnamefont
  {Dimopoulos}}, \ and\ \bibinfo {author} {\bibfnamefont {K.}~\bibnamefont
  {Van~Tilburg}},\ }\href {\doibase 10.1103/PhysRevLett.116.031102} {\bibfield
  {journal} {\bibinfo  {journal} {Phys. Rev. Lett.}\ }\textbf {\bibinfo
  {volume} {116}},\ \bibinfo {pages} {031102} (\bibinfo {year} {2016})},\
  \Eprint {http://arxiv.org/abs/1508.01798} {arXiv:1508.01798 [hep-ph]}
  \BibitemShut {NoStop}%
\bibitem [{\citenamefont {Graham}\ \emph
  {et~al.}(2015{\natexlab{b}})\citenamefont {Graham}, \citenamefont {Kaplan},
  \citenamefont {Mardon}, \citenamefont {Rajendran},\ and\ \citenamefont
  {Terrano}}]{Graham:2015ifn}%
  \BibitemOpen
  \bibfield  {author} {\bibinfo {author} {\bibfnamefont {P.~W.}\ \bibnamefont
  {Graham}}, \bibinfo {author} {\bibfnamefont {D.~E.}\ \bibnamefont {Kaplan}},
  \bibinfo {author} {\bibfnamefont {J.}~\bibnamefont {Mardon}}, \bibinfo
  {author} {\bibfnamefont {S.}~\bibnamefont {Rajendran}}, \ and\ \bibinfo
  {author} {\bibfnamefont {W.~A.}\ \bibnamefont {Terrano}},\ }\href@noop {} {\
  (\bibinfo {year} {2015}{\natexlab{b}})},\ \Eprint
  {http://arxiv.org/abs/1512.06165} {arXiv:1512.06165 [hep-ph]} \BibitemShut
  {NoStop}%
\bibitem [{\citenamefont {Graham}\ \emph {et~al.}(2013)\citenamefont {Graham},
  \citenamefont {Hogan}, \citenamefont {Kasevich},\ and\ \citenamefont
  {Rajendran}}]{Graham:2012sy}%
  \BibitemOpen
  \bibfield  {author} {\bibinfo {author} {\bibfnamefont {P.~W.}\ \bibnamefont
  {Graham}}, \bibinfo {author} {\bibfnamefont {J.~M.}\ \bibnamefont {Hogan}},
  \bibinfo {author} {\bibfnamefont {M.~A.}\ \bibnamefont {Kasevich}}, \ and\
  \bibinfo {author} {\bibfnamefont {S.}~\bibnamefont {Rajendran}},\ }\href
  {\doibase 10.1103/PhysRevLett.110.171102} {\bibfield  {journal} {\bibinfo
  {journal} {Phys. Rev. Lett.}\ }\textbf {\bibinfo {volume} {110}},\ \bibinfo
  {pages} {171102} (\bibinfo {year} {2013})},\ \Eprint
  {http://arxiv.org/abs/1206.0818} {arXiv:1206.0818 [quant-ph]} \BibitemShut
  {NoStop}%
\bibitem [{\citenamefont {Damour}\ and\ \citenamefont
  {Donoghue}(2010{\natexlab{a}})}]{Damour:2010rm}%
  \BibitemOpen
  \bibfield  {author} {\bibinfo {author} {\bibfnamefont {T.}~\bibnamefont
  {Damour}}\ and\ \bibinfo {author} {\bibfnamefont {J.~F.}\ \bibnamefont
  {Donoghue}},\ }\href {\doibase 10.1088/0264-9381/27/20/202001} {\bibfield
  {journal} {\bibinfo  {journal} {Class. Quant. Grav.}\ }\textbf {\bibinfo
  {volume} {27}},\ \bibinfo {pages} {202001} (\bibinfo {year}
  {2010}{\natexlab{a}})},\ \Eprint {http://arxiv.org/abs/1007.2790}
  {arXiv:1007.2790 [gr-qc]} \BibitemShut {NoStop}%
\bibitem [{\citenamefont {Damour}\ and\ \citenamefont
  {Donoghue}(2010{\natexlab{b}})}]{Damour:2010rp}%
  \BibitemOpen
  \bibfield  {author} {\bibinfo {author} {\bibfnamefont {T.}~\bibnamefont
  {Damour}}\ and\ \bibinfo {author} {\bibfnamefont {J.~F.}\ \bibnamefont
  {Donoghue}},\ }\href {\doibase 10.1103/PhysRevD.82.084033} {\bibfield
  {journal} {\bibinfo  {journal} {Phys.Rev.}\ }\textbf {\bibinfo {volume}
  {D82}},\ \bibinfo {pages} {084033} (\bibinfo {year} {2010}{\natexlab{b}})},\
  \Eprint {http://arxiv.org/abs/1007.2792} {arXiv:1007.2792 [gr-qc]}
  \BibitemShut {NoStop}%
\bibitem [{\citenamefont {Olive}\ and\ \citenamefont
  {Pospelov}(2008)}]{PhysRevD.77.043524}%
  \BibitemOpen
  \bibfield  {author} {\bibinfo {author} {\bibfnamefont {K.~A.}\ \bibnamefont
  {Olive}}\ and\ \bibinfo {author} {\bibfnamefont {M.}~\bibnamefont
  {Pospelov}},\ }\href {\doibase 10.1103/PhysRevD.77.043524} {\bibfield
  {journal} {\bibinfo  {journal} {Phys. Rev. D}\ }\textbf {\bibinfo {volume}
  {77}},\ \bibinfo {pages} {043524} (\bibinfo {year} {2008})}\BibitemShut
  {NoStop}%
\bibitem [{\citenamefont {Stadnik}\ and\ \citenamefont
  {Flambaum}(2015)}]{PhysRevLett.114.161301}%
  \BibitemOpen
  \bibfield  {author} {\bibinfo {author} {\bibfnamefont {Y.~V.}\ \bibnamefont
  {Stadnik}}\ and\ \bibinfo {author} {\bibfnamefont {V.~V.}\ \bibnamefont
  {Flambaum}},\ }\href {\doibase 10.1103/PhysRevLett.114.161301} {\bibfield
  {journal} {\bibinfo  {journal} {Phys. Rev. Lett.}\ }\textbf {\bibinfo
  {volume} {114}},\ \bibinfo {pages} {161301} (\bibinfo {year}
  {2015})}\BibitemShut {NoStop}%
\bibitem [{\citenamefont {Angstmann}\ \emph {et~al.}(2004)\citenamefont
  {Angstmann}, \citenamefont {Dzuba},\ and\ \citenamefont
  {Flambaum}}]{PhysRevA.70.014102}%
  \BibitemOpen
  \bibfield  {author} {\bibinfo {author} {\bibfnamefont {E.~J.}\ \bibnamefont
  {Angstmann}}, \bibinfo {author} {\bibfnamefont {V.~A.}\ \bibnamefont
  {Dzuba}}, \ and\ \bibinfo {author} {\bibfnamefont {V.~V.}\ \bibnamefont
  {Flambaum}},\ }\href {\doibase 10.1103/PhysRevA.70.014102} {\bibfield
  {journal} {\bibinfo  {journal} {Phys. Rev. A}\ }\textbf {\bibinfo {volume}
  {70}},\ \bibinfo {pages} {014102} (\bibinfo {year} {2004})}\BibitemShut
  {NoStop}%
\bibitem [{\citenamefont {Geraci}\ and\ \citenamefont
  {Derevianko}(2016)}]{Geraci:2016fva}%
  \BibitemOpen
  \bibfield  {author} {\bibinfo {author} {\bibfnamefont {A.~A.}\ \bibnamefont
  {Geraci}}\ and\ \bibinfo {author} {\bibfnamefont {A.}~\bibnamefont
  {Derevianko}},\ }\href@noop {} {\  (\bibinfo {year} {2016})},\ \Eprint
  {http://arxiv.org/abs/1605.04048} {arXiv:1605.04048 [physics.atom-ph]}
  \BibitemShut {NoStop}%
\bibitem [{\citenamefont {Hosten}\ \emph {et~al.}(2016)\citenamefont {Hosten},
  \citenamefont {Engelsen}, \citenamefont {Krishnakumar},\ and\ \citenamefont
  {Kasevich}}]{hosten2016measurement}%
  \BibitemOpen
  \bibfield  {author} {\bibinfo {author} {\bibfnamefont {O.}~\bibnamefont
  {Hosten}}, \bibinfo {author} {\bibfnamefont {N.~J.}\ \bibnamefont
  {Engelsen}}, \bibinfo {author} {\bibfnamefont {R.}~\bibnamefont
  {Krishnakumar}}, \ and\ \bibinfo {author} {\bibfnamefont {M.~A.}\
  \bibnamefont {Kasevich}},\ }\href {\doibase 10.1038/nature16176} {\bibfield
  {journal} {\bibinfo  {journal} {Nature}\ }\textbf {\bibinfo {volume} {529}},\
  \bibinfo {pages} {505} (\bibinfo {year} {2016})}\BibitemShut {NoStop}%
\bibitem [{\citenamefont {Hogan}\ and\ \citenamefont
  {Kasevich}(2015)}]{Hogan:2015xla}%
  \BibitemOpen
  \bibfield  {author} {\bibinfo {author} {\bibfnamefont {J.~M.}\ \bibnamefont
  {Hogan}}\ and\ \bibinfo {author} {\bibfnamefont {M.~A.}\ \bibnamefont
  {Kasevich}},\ }\href@noop {} {\  (\bibinfo {year} {2015})},\ \Eprint
  {http://arxiv.org/abs/1501.06797} {arXiv:1501.06797 [physics.atom-ph]}
  \BibitemShut {NoStop}%
\bibitem [{\citenamefont {Dimopoulos}\ \emph {et~al.}(2009)\citenamefont
  {Dimopoulos}, \citenamefont {Graham}, \citenamefont {Hogan}, \citenamefont
  {Kasevich},\ and\ \citenamefont {Rajendran}}]{Dimopoulos:2007cj}%
  \BibitemOpen
  \bibfield  {author} {\bibinfo {author} {\bibfnamefont {S.}~\bibnamefont
  {Dimopoulos}}, \bibinfo {author} {\bibfnamefont {P.~W.}\ \bibnamefont
  {Graham}}, \bibinfo {author} {\bibfnamefont {J.~M.}\ \bibnamefont {Hogan}},
  \bibinfo {author} {\bibfnamefont {M.~A.}\ \bibnamefont {Kasevich}}, \ and\
  \bibinfo {author} {\bibfnamefont {S.}~\bibnamefont {Rajendran}},\ }\href
  {\doibase 10.1016/j.physletb.2009.06.011} {\bibfield  {journal} {\bibinfo
  {journal} {Phys. Lett.}\ }\textbf {\bibinfo {volume} {B678}},\ \bibinfo
  {pages} {37} (\bibinfo {year} {2009})},\ \Eprint
  {http://arxiv.org/abs/0712.1250} {arXiv:0712.1250 [gr-qc]} \BibitemShut
  {NoStop}%
\bibitem [{\citenamefont {Dimopoulos}\ \emph {et~al.}(2008)\citenamefont
  {Dimopoulos}, \citenamefont {Graham}, \citenamefont {Hogan}, \citenamefont
  {Kasevich},\ and\ \citenamefont {Rajendran}}]{Dimopoulos:2008sv}%
  \BibitemOpen
  \bibfield  {author} {\bibinfo {author} {\bibfnamefont {S.}~\bibnamefont
  {Dimopoulos}}, \bibinfo {author} {\bibfnamefont {P.~W.}\ \bibnamefont
  {Graham}}, \bibinfo {author} {\bibfnamefont {J.~M.}\ \bibnamefont {Hogan}},
  \bibinfo {author} {\bibfnamefont {M.~A.}\ \bibnamefont {Kasevich}}, \ and\
  \bibinfo {author} {\bibfnamefont {S.}~\bibnamefont {Rajendran}},\ }\href
  {\doibase 10.1103/PhysRevD.78.122002} {\bibfield  {journal} {\bibinfo
  {journal} {Phys. Rev.}\ }\textbf {\bibinfo {volume} {D78}},\ \bibinfo {pages}
  {122002} (\bibinfo {year} {2008})},\ \Eprint {http://arxiv.org/abs/0806.2125}
  {arXiv:0806.2125 [gr-qc]} \BibitemShut {NoStop}%
\bibitem [{\citenamefont {Schlamminger}\ \emph {et~al.}(2008)\citenamefont
  {Schlamminger}, \citenamefont {Choi}, \citenamefont {Wagner}, \citenamefont
  {Gundlach},\ and\ \citenamefont {Adelberger}}]{Schlamminger:2007ht}%
  \BibitemOpen
  \bibfield  {author} {\bibinfo {author} {\bibfnamefont {S.}~\bibnamefont
  {Schlamminger}}, \bibinfo {author} {\bibfnamefont {K.-Y.}\ \bibnamefont
  {Choi}}, \bibinfo {author} {\bibfnamefont {T.~A.}\ \bibnamefont {Wagner}},
  \bibinfo {author} {\bibfnamefont {J.~H.}\ \bibnamefont {Gundlach}}, \ and\
  \bibinfo {author} {\bibfnamefont {E.~G.}\ \bibnamefont {Adelberger}},\ }\href
  {\doibase 10.1103/PhysRevLett.100.041101} {\bibfield  {journal} {\bibinfo
  {journal} {Phys.Rev.Lett.}\ }\textbf {\bibinfo {volume} {100}},\ \bibinfo
  {pages} {041101} (\bibinfo {year} {2008})},\ \Eprint
  {http://arxiv.org/abs/0712.0607} {arXiv:0712.0607 [gr-qc]} \BibitemShut
  {NoStop}%
\bibitem [{\citenamefont {Wagner}\ \emph {et~al.}(2012)\citenamefont {Wagner},
  \citenamefont {Schlamminger}, \citenamefont {Gundlach},\ and\ \citenamefont
  {Adelberger}}]{Wagner:2012ui}%
  \BibitemOpen
  \bibfield  {author} {\bibinfo {author} {\bibfnamefont {T.~A.}\ \bibnamefont
  {Wagner}}, \bibinfo {author} {\bibfnamefont {S.}~\bibnamefont
  {Schlamminger}}, \bibinfo {author} {\bibfnamefont {J.~H.}\ \bibnamefont
  {Gundlach}}, \ and\ \bibinfo {author} {\bibfnamefont {E.~G.}\ \bibnamefont
  {Adelberger}},\ }\href {\doibase 10.1088/0264-9381/29/18/184002} {\bibfield
  {journal} {\bibinfo  {journal} {Class. Quant. Grav.}\ }\textbf {\bibinfo
  {volume} {29}},\ \bibinfo {pages} {184002} (\bibinfo {year} {2012})},\
  \Eprint {http://arxiv.org/abs/1207.2442} {arXiv:1207.2442 [gr-qc]}
  \BibitemShut {NoStop}%
\bibitem [{\citenamefont {Adelberger}\ \emph {et~al.}(2003)\citenamefont
  {Adelberger}, \citenamefont {Heckel},\ and\ \citenamefont
  {Nelson}}]{Adelberger:2003zx}%
  \BibitemOpen
  \bibfield  {author} {\bibinfo {author} {\bibfnamefont {E.}~\bibnamefont
  {Adelberger}}, \bibinfo {author} {\bibfnamefont {B.~R.}\ \bibnamefont
  {Heckel}}, \ and\ \bibinfo {author} {\bibfnamefont {A.}~\bibnamefont
  {Nelson}},\ }\href {\doibase 10.1146/annurev.nucl.53.041002.110503}
  {\bibfield  {journal} {\bibinfo  {journal} {Ann.Rev.Nucl.Part.Sci.}\ }\textbf
  {\bibinfo {volume} {53}},\ \bibinfo {pages} {77} (\bibinfo {year} {2003})},\
  \Eprint {http://arxiv.org/abs/hep-ph/0307284} {arXiv:hep-ph/0307284 [hep-ph]}
  \BibitemShut {NoStop}%
\bibitem [{\citenamefont {Weiss}\ and\ \citenamefont
  {Block}(1965)}]{JGR:JGR4118}%
  \BibitemOpen
  \bibfield  {author} {\bibinfo {author} {\bibfnamefont {R.}~\bibnamefont
  {Weiss}}\ and\ \bibinfo {author} {\bibfnamefont {B.}~\bibnamefont {Block}},\
  }\href {\doibase 10.1029/JZ070i022p05615} {\bibfield  {journal} {\bibinfo
  {journal} {Journal of Geophysical Research}\ }\textbf {\bibinfo {volume}
  {70}},\ \bibinfo {pages} {5615} (\bibinfo {year} {1965})}\BibitemShut
  {NoStop}%
\bibitem [{\citenamefont {Baggio}\ \emph {et~al.}(2005)\citenamefont {Baggio},
  \citenamefont {Bignotto}, \citenamefont {Bonaldi}, \citenamefont {Cerdonio},
  \citenamefont {Conti}, \citenamefont {Falferi}, \citenamefont {Liguori},
  \citenamefont {Marin}, \citenamefont {Mezzena}, \citenamefont {Ortolan},
  \citenamefont {Poggi}, \citenamefont {Prodi}, \citenamefont {Salemi},
  \citenamefont {Soranzo}, \citenamefont {Taffarello}, \citenamefont
  {Vedovato}, \citenamefont {Vinante}, \citenamefont {Vitale},\ and\
  \citenamefont {Zendri}}]{auriga1}%
  \BibitemOpen
  \bibfield  {author} {\bibinfo {author} {\bibfnamefont {L.}~\bibnamefont
  {Baggio}}, \bibinfo {author} {\bibfnamefont {M.}~\bibnamefont {Bignotto}},
  \bibinfo {author} {\bibfnamefont {M.}~\bibnamefont {Bonaldi}}, \bibinfo
  {author} {\bibfnamefont {M.}~\bibnamefont {Cerdonio}}, \bibinfo {author}
  {\bibfnamefont {L.}~\bibnamefont {Conti}}, \bibinfo {author} {\bibfnamefont
  {P.}~\bibnamefont {Falferi}}, \bibinfo {author} {\bibfnamefont
  {N.}~\bibnamefont {Liguori}}, \bibinfo {author} {\bibfnamefont
  {A.}~\bibnamefont {Marin}}, \bibinfo {author} {\bibfnamefont
  {R.}~\bibnamefont {Mezzena}},  \emph {et~al.},\ }\href {\doibase
  10.1103/PhysRevLett.94.241101} {\bibfield  {journal} {\bibinfo  {journal}
  {Phys. Rev. Lett.}\ }\textbf {\bibinfo {volume} {94}},\ \bibinfo {pages}
  {241101} (\bibinfo {year} {2005})}\BibitemShut {NoStop}%
\bibitem [{\citenamefont {Graham}\ \emph {et~al.}(2016)\citenamefont {Graham},
  \citenamefont {Hogan}, \citenamefont {Kasevich},\ and\ \citenamefont
  {Rajendran}}]{Graham:2016plp}%
  \BibitemOpen
  \bibfield  {author} {\bibinfo {author} {\bibfnamefont {P.~W.}\ \bibnamefont
  {Graham}}, \bibinfo {author} {\bibfnamefont {J.~M.}\ \bibnamefont {Hogan}},
  \bibinfo {author} {\bibfnamefont {M.~A.}\ \bibnamefont {Kasevich}}, \ and\
  \bibinfo {author} {\bibfnamefont {S.}~\bibnamefont {Rajendran}},\ }\href@noop
  {} {\  (\bibinfo {year} {2016})},\ \Eprint {http://arxiv.org/abs/1606.01860}
  {arXiv:1606.01860 [physics.atom-ph]} \BibitemShut {NoStop}%
\bibitem [{\citenamefont {Derevianko}\ and\ \citenamefont
  {Pospelov}(2014)}]{Derevianko:2013oaa}%
  \BibitemOpen
  \bibfield  {author} {\bibinfo {author} {\bibfnamefont {A.}~\bibnamefont
  {Derevianko}}\ and\ \bibinfo {author} {\bibfnamefont {M.}~\bibnamefont
  {Pospelov}},\ }\href {\doibase 10.1038/nphys3137} {\bibfield  {journal}
  {\bibinfo  {journal} {Nature Phys.}\ }\textbf {\bibinfo {volume} {10}},\
  \bibinfo {pages} {933} (\bibinfo {year} {2014})},\ \Eprint
  {http://arxiv.org/abs/1311.1244} {arXiv:1311.1244 [physics.atom-ph]}
  \BibitemShut {NoStop}%
\bibitem [{\citenamefont {Collaboration}(2015)}]{0264-9381-32-7-074001}%
  \BibitemOpen
  \bibfield  {author} {\bibinfo {author} {\bibfnamefont {T.~L.~S.}\
  \bibnamefont {Collaboration}},\ }\href
  {http://stacks.iop.org/0264-9381/32/i=7/a=074001} {\bibfield  {journal}
  {\bibinfo  {journal} {Classical and Quantum Gravity}\ }\textbf {\bibinfo
  {volume} {32}},\ \bibinfo {pages} {074001} (\bibinfo {year}
  {2015})}\BibitemShut {NoStop}%
\bibitem [{\citenamefont {{Wcislo}}\ \emph {et~al.}(2016)\citenamefont
  {{Wcislo}}, \citenamefont {{Morzynski}}, \citenamefont {{Bober}},
  \citenamefont {{Cygan}}, \citenamefont {{Lisak}}, \citenamefont {{Ciurylo}},\
  and\ \citenamefont {{Zawada}}}]{2016arXiv160505763W}%
  \BibitemOpen
  \bibfield  {author} {\bibinfo {author} {\bibfnamefont {P.}~\bibnamefont
  {{Wcislo}}}, \bibinfo {author} {\bibfnamefont {P.}~\bibnamefont
  {{Morzynski}}}, \bibinfo {author} {\bibfnamefont {M.}~\bibnamefont
  {{Bober}}}, \bibinfo {author} {\bibfnamefont {A.}~\bibnamefont {{Cygan}}},
  \bibinfo {author} {\bibfnamefont {D.}~\bibnamefont {{Lisak}}}, \bibinfo
  {author} {\bibfnamefont {R.}~\bibnamefont {{Ciurylo}}}, \ and\ \bibinfo
  {author} {\bibfnamefont {M.}~\bibnamefont {{Zawada}}},\ }\href@noop {}
  {\bibfield  {journal} {\bibinfo  {journal} {ArXiv e-prints}\ } (\bibinfo
  {year} {2016})},\ \Eprint {http://arxiv.org/abs/1605.05763} {arXiv:1605.05763
  [physics.atom-ph]} \BibitemShut {NoStop}%
\bibitem [{\citenamefont {Amaro-Seoane}\ \emph {et~al.}(2012)\citenamefont
  {Amaro-Seoane}, \citenamefont {Aoudia}, \citenamefont {Babak}, \citenamefont
  {Binetruy}, \citenamefont {Berti} \emph {et~al.}}]{LISA}%
  \BibitemOpen
  \bibfield  {author} {\bibinfo {author} {\bibfnamefont {P.}~\bibnamefont
  {Amaro-Seoane}}, \bibinfo {author} {\bibfnamefont {S.}~\bibnamefont
  {Aoudia}}, \bibinfo {author} {\bibfnamefont {S.}~\bibnamefont {Babak}},
  \bibinfo {author} {\bibfnamefont {P.}~\bibnamefont {Binetruy}}, \bibinfo
  {author} {\bibfnamefont {E.}~\bibnamefont {Berti}},  \emph {et~al.},\ }\href
  {\doibase 10.1088/0264-9381/29/12/124016} {\emph {\bibinfo {title}
  {{Low-frequency gravitational-wave science with eLISA/NGO}}}},\ \bibinfo
  {type} {Tech. Rep.}\ (\bibinfo {year} {2012})\ \Eprint
  {http://arxiv.org/abs/1202.0839} {arXiv:1202.0839 [gr-qc]} \BibitemShut
  {NoStop}%
\bibitem [{\citenamefont {Hogan}\ \emph {et~al.}(2011)\citenamefont {Hogan}
  \emph {et~al.}}]{Hogan:2010fz}%
  \BibitemOpen
  \bibfield  {author} {\bibinfo {author} {\bibfnamefont {J.~M.}\ \bibnamefont
  {Hogan}} \emph {et~al.},\ }\href {\doibase 10.1007/s10714-011-1182-x}
  {\bibfield  {journal} {\bibinfo  {journal} {Gen. Rel. Grav.}\ }\textbf
  {\bibinfo {volume} {43}},\ \bibinfo {pages} {1953} (\bibinfo {year}
  {2011})},\ \Eprint {http://arxiv.org/abs/1009.2702} {arXiv:1009.2702
  [physics.atom-ph]} \BibitemShut {NoStop}%
\end{thebibliography}%

\end{document}